\documentclass[lettersize,journal]{IEEEtran}
\usepackage{amsmath,amssymb,amsfonts,amsthm}
\usepackage{graphicx,booktabs,multirow,array}
\graphicspath{{figures/}}
\usepackage{float}
\usepackage{cite}
\usepackage{algorithm}
\usepackage{algpseudocode}
\usepackage[table]{xcolor}
\usepackage[caption=false,font=normalsize,labelfont=sf,textfont=sf]{subfig}
\usepackage{textcomp,stfloats,url,verbatim}
\usepackage[hidelinks]{hyperref}
\hypersetup{pdftitle={LiBRA: Detection-Aware Image Watermark Removal via Bidirectional Latent Optimization},
pdfsubject={TDSC manuscript}
}
\newtheorem{proposition}{Proposition}

\newcommand{\prg}[1]{\vskip 2pt \noindent \textbf{#1}.}

\usepackage{array}

\begin{document}
\title{LiBRA: Detection-Aware Image Watermark Removal via Bidirectional Latent Optimization}

\author{%
Saibo~Ye$^{\spadesuit,\dagger}$,
Huajie~Chen$^{\spadesuit,\diamondsuit,\dagger}$,
Xin~Guo$^{\dagger}$,
Le~Yang$^{\diamondsuit,\ddagger}$,
Chi~Liu$^{\dagger}$,
Xiangyu~Hu$^{\S}$,
Jingjing~Guo$^{\P}$,
and~Tianqing~Zhu$^{\dagger}$%
\thanks{% 
$^\spadesuit$Equal Contribution, $^\diamondsuit$Corresponding Author. $^\dagger$Faculty of Data Science, City University of Macau, Macao SAR, China. $^\ddagger$School of Information and Communications Engineering, Xi'an Jiaotong University, Xi'an 710049, China. $^{\S}$School of Information and Software Engineering, University of Electronic Science and Technology of China, Chengdu 610054, China. $^{\P}$School of Cyber Engineering, Xidian University, Xi'an 710071, China. E-mail: \{d24090127012, hjchen, h23090104001, chiliu, tqzhu\}@cityu.edu.mo; yangle15@xjtu.edu.cn; huxiangyu@uestc.edu.cn; jjguo@xidian.edu.cn.}%
}

\maketitle

\begin{abstract}
Digital watermarking supports source attribution for AI-generated images, but its reliability depends on resistance to removal attacks. Some attacks attempt to remove watermarks by forcing the decoded watermark to differ from the original. However, this can produce an inverted watermark that remains detectable, causing removal to fail, while further attempts to alter the watermark may unnecessarily degrade image quality. To address these limitations, we present LiBRA (Latent In-band Bidirectional Removal Attack), which aims to make watermarks undetectable while preserving image quality. Instead of continually pushing the watermark toward inversion, LiBRA adjusts the image to conceal the watermark without encouraging further changes that could degrade image quality. Some attacks keep pushing decoded bits away from the original watermark, even when further changes preserve detectability and damage image quality. With access to the watermark key and decoder, LiBRA makes bounded changes in a public autoencoder's latent space. Unlike inversion-driven objectives that cannot correct excessive inversion, LiBRA guides average decoding confidence toward random guessing from either direction. This helps avoid an inverted but detectable watermark. Leaving individual bits flexible allows image-quality constraints to favor less damaging changes, while an optional frequency-guided mask limits their location. We verify removal using an exact two-sided binomial test rather than assuming the confidence target guarantees success. Experiments on four released watermarking systems show that LiBRA substantially reduces two-sided detection rates while improving image fidelity over the inversion baseline. An iteration study on Stable Signature shows that extending optimization from $25$ to $100$ iterations further degrades the inversion baseline's image quality, while LiBRA maintains stable detection rates and image fidelity.
\end{abstract}

\section{Introduction}\label{sec:intro}
\IEEEPARstart{T}{he} rapid development of artificial intelligence-generated content (AIGC) has made high-quality synthetic images increasingly accessible. Generative adversarial networks, diffusion models, and latent diffusion models have advanced visual realism, supporting content creation, digital design, and media production~\cite{goodfellow2014gan,ho2020ddpm,rombach2022ldm}. These capabilities also complicate authenticity verification and source attribution, creating risks of misleading dissemination and unauthorized use~\cite{verdoliva2020media}. Digital watermarking supports provenance by embedding recoverable identifiers into images. Research has progressed from conventional signal-based embedding to learned encoder--decoder systems and generation-integrated watermarking~\cite{cox1997,zhu2018hidden,fernandez2023stable,zhao2025sok}. Despite their different embedding mechanisms, these methods depend on reliable detection after image processing and deliberate removal attempts.

Alongside watermarking, removal attacks have expanded beyond conventional transformations to exploit decoder feedback, spectral structures, and generative models. Pixel-domain attacks directly optimize image perturbations, as exemplified by WEvade and its WEvade-W-II variant, which uses random target watermarks to address double-tail detection~\cite{jiang2023wevade}. Spectral attacks disrupt structures associated with watermark carriers; UnMarker does so without detector feedback~\cite{kassis2025unmarker}. Regeneration attacks reconstruct images through learned compression or diffusion models to weaken watermarks while preserving visual content~\cite{zhao2024invisible,liu2025ctrlregen}. Latent-space optimization offers another route, modifying compact autoencoder representations before decoding them into images. Related research also explores decision-boundary information for watermark removal~\cite{lee2025boundary}. Although their access assumptions and distortion mechanisms differ, these approaches must balance detection evasion with image fidelity.

\begin{figure}[t!]
    \centering
    \includegraphics[width=\linewidth]{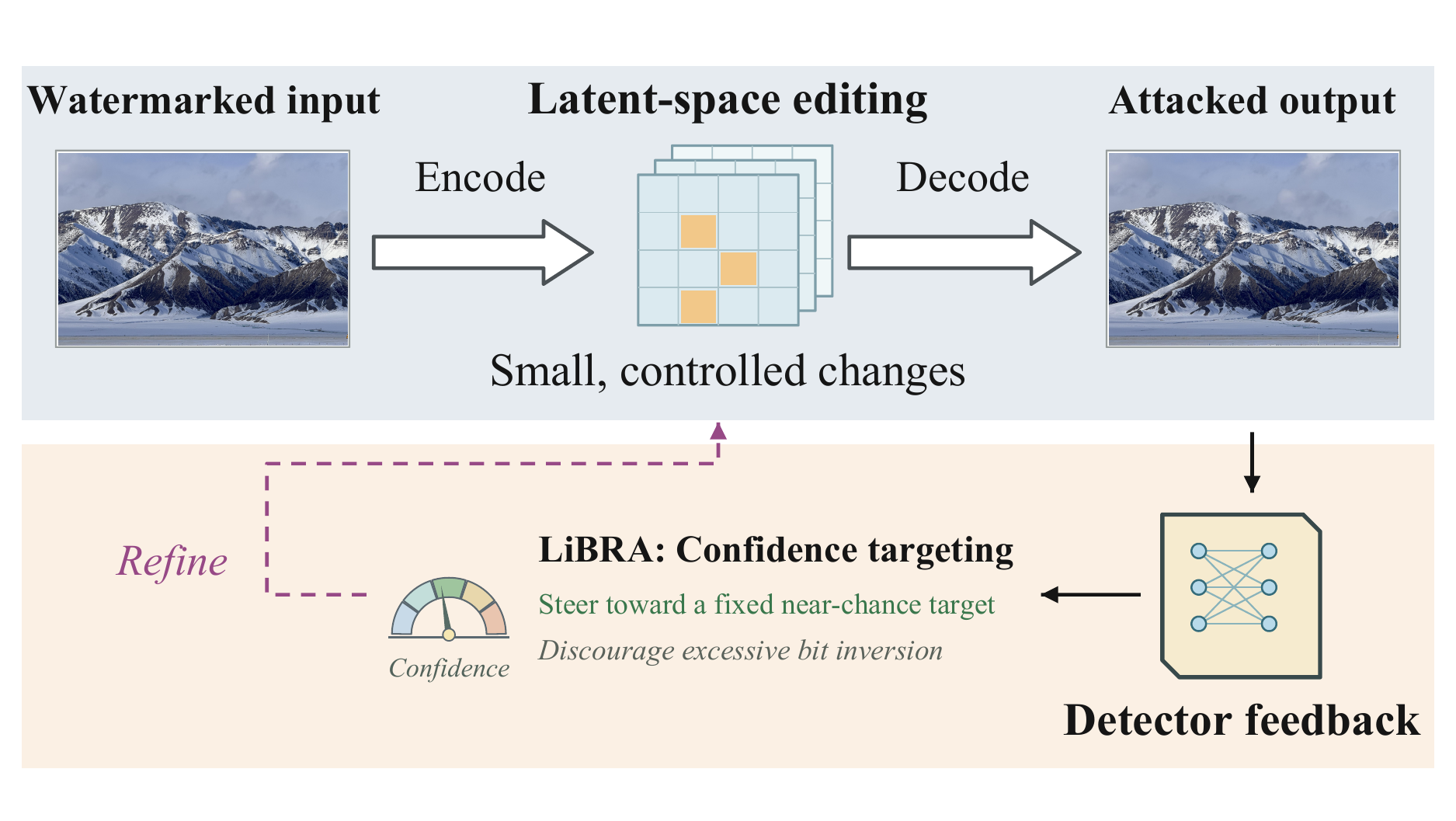}
    \caption{Overview of LiBRA. A watermarked image is encoded
    into latent space and iteratively refined using feedback
    from a frozen watermark detector. LiBRA steers the mean
    soft bit-matching confidence toward a fixed near-chance
    target to discourage excessive bit inversion, while
    fidelity regularization limits visual changes. The refined
    latent is decoded into the output image. The dashed arrow
    denotes iterative refinement. Images are illustrative.}
    \label{fig:overview}
\end{figure}

Three issues remain. First, low bit accuracy does not necessarily indicate removal: pushing decoded bits away from the key can produce an inverted watermark that remains detectable under a two-sided test. Second, objectives that reward continued inversion may introduce distortion after entering the non-detection region, while simple per-bit stopping thresholds can admit fully inverted results. Third, attacks optimize soft confidences, whereas detection depends on discrete bit matches, and these quantities can disagree. We examine whether a bidirectional objective on average decoding confidence can counter excessive inversion while preserving image quality during latent-space optimization. Evaluation must examine image-level detection outcomes, since average bit accuracy near chance can conceal removal failures.

\prg{Our Work}
We propose LiBRA (Latent In-band Bidirectional Removal Attack), a detection-aware watermark removal method operating in latent space. We consider a white-box setting in which the adversary can access the watermark decoder and its gradients and knows or can estimate the watermark key. Our goal is to evade a two-sided statistical test while preserving the visual content of the input image. Given a watermarked image, we first encode it using a public, non-watermarked autoencoder, optimize a bounded perturbation in its latent representation, and decode the modified representation into an attacked image. This allows the attack to operate in a compact space without accessing the provider's watermark-generating components. Latent and image-level fidelity penalties constrain changes to the representation and reconstructed image. An optional frequency-guided mask further restricts perturbations to regions selected from the image content, providing a spatial prior for controlling visual distortion.

The central component is a symmetric loss that guides the mean per-bit confidence toward a prescribed near-chance target. When the mean exceeds the target, the loss encourages it to decrease; when it falls below the target, the loss encourages it to increase. The objective therefore counteracts excessive inversion while leaving individual bit configurations unspecified. We optimize this loss jointly with the fidelity terms, allowing image preservation to influence how the target is approached. LiBRA uses a single optimization procedure without an outer perturbation-budget search. We evaluate the resulting images with the two-sided test and examine the gap between soft confidence and hard bit accuracy. Comparisons across the evaluated watermarking systems and ablations of supporting components assess whether the objective improves the balance between evasion and fidelity.

Our main contributions are as follows:
\begin{itemize}
\item \textbf{Analysis of objective--detector mismatch.} We explain why confidence-minimizing objectives and simple per-bit thresholding rules can still admit fully inverted watermarks, and distinguish soft-confidence optimization from hard-bit detection.
\item \textbf{A detection-aware aggregate objective in latent space.} We study a symmetric mean-confidence penalty combined with bounded latent optimization, fidelity constraints, and an optional spatial prior, examining its effect on watermark evasion and image quality.
\item \textbf{Cross-system evaluation and mechanism analysis.} We assess four watermarking systems through direct comparisons with WEvade-W-II, component ablations, surrogate-gap measurements, and adaptive evaluation, establishing the approach's empirical benefits and scope under the stated threat model.
\end{itemize}

\section{Preliminaries}\label{sec:prelim}
\autoref{tab:notations} summarizes the principal notation used
in the subsequent analysis, method, and experiments.
\begin{table}[t!]
\centering
\caption{Main notation.}
\label{tab:notations}
\footnotesize
\renewcommand{\arraystretch}{1.1}
\setlength{\tabcolsep}{4pt}
\rowcolors{2}{gray!10}{white}
\begin{tabular}{@{}>{\centering\arraybackslash}p{0.29\columnwidth}p{0.66\columnwidth}@{}}
\toprule
\rowcolor{white}
\textbf{Notation} & \textbf{Definition} \\
\midrule
$x_0$, $x'$ &
Watermarked input and attacked image. \\

$E$, $D$ &
Public autoencoder encoder and decoder. \\

$D_m$ &
Watermark decoder. \\

$w$, $N$ &
Watermark key and its bit length. \\

$L_w$, $\delta$ &
Input latent representation and latent perturbation. \\

$M^{\mathrm{s}}$ &
Frequency-guided spatial mask. \\

$a(x')$ &
Fraction of decoded bits matching the key. \\

$\bar c$, $a^{\star}$ &
Mean soft confidence of agreement with the key and its target value. \\

$\mathcal A_{N,\alpha}$ &
Accepted bit accuracies under the two-sided test
at significance level $\alpha$. \\

$L_{\mathrm{bd2}}$, $L_{\mathrm{fid}}$ &
Proposed watermark loss and fidelity loss. \\
\bottomrule
\end{tabular}
\end{table}

\subsection{Notation}
We write $x_0\in\mathbb{R}^{3\times H\times W}$ for a watermarked image and $x'$ for its attacked counterpart. The provider's secret key is $w\in\{-1,+1\}^{N}$ and its detector is $D_m$, whose soft output on an image is $d=D_m(x')\in\mathbb{R}^{N}$. We write $s_i=d_i w_i$ for the per-bit margin, so that $s_i>0$ means bit $i$ is decoded correctly, and $c_i=\sigma(s_i)$ for the corresponding confidence, where $\sigma$ is the logistic sigmoid. The mean confidence is $\bar c=N^{-1}\sum_{i=1}^{N}c_i$, whereas hard bit accuracy $a(x')$ is the fraction of decoded bits matching the key. Vectors on the latent grid are written with a capital $L$; $\odot$ denotes the Hadamard product and $\lVert\cdot\rVert_0$ counts nonzeros.

\subsection{Latent Autoencoding}

A variational autoencoder consists of an encoder $E$ and a decoder $D$~\cite{kingma2014vae}. The encoder maps an image to a lower-dimensional representation,
\[
L=E(x)\in\mathbb{R}^{C\times(H/f)\times(W/f)},
\]
where $C$ is the latent channel count and $f$ is the spatial downsampling factor. The decoder reconstructs an image from this representation. Latent diffusion models use such representations for image generation~\cite{rombach2022ldm}.

For $C=4$ and $f=8$, a $512\times512$ RGB image corresponds to a $4\times64\times64$ latent representation, reducing the number of optimization variables by a factor of $48$. A differentiable decoder allows gradients computed on the reconstructed image to propagate to the latent representation. However, autoencoder reconstruction is lossy, and small latent changes do not by themselves guarantee small perceptual changes.

\subsection{Discrete Wavelet Decomposition}

A one-level two-dimensional discrete wavelet transform separates an image into an approximation subband $Y_L$ and three detail subbands $Y_{LH}$, $Y_{HL}$, and $Y_{HH}$~\cite{mallat1989}. The approximation subband represents coarse image structure, while the detail subbands describe directional changes at a finer scale. Each subband has half the height and width of the input.

The Haar transform computes these components using local sums and differences. Its spatial localization allows subband magnitudes to describe where image structure and texture occur. Such information can support a spatial prior, but does not identify where a watermark is embedded. For an orthonormal transform with compatible boundary handling, the sum of squared errors over all subbands equals the squared error in the image domain.

\subsection{Watermark Detection as a Hypothesis Test}\label{sec:rw_det}

A watermark decoder produces bit predictions, which the provider converts into a detection decision. Under the unwatermarked null hypothesis $\mathcal{H}_0$, decoded bits are assumed to agree with a fixed key independently with probability $1/2$. The random number of matching bits therefore follows
\[
K\sim\mathrm{Binomial}(N,1/2).
\]
For an observed image, the matching count is $k=Na(x')$.

A one-sided test detects unusually high agreement with the key. A two-sided test also detects unusually low agreement, since a nearly inverted key is unlikely under the same null model. Consequently, both near-perfect agreement and near-complete inversion can provide evidence of a watermark~\cite{jiang2023wevade}.

At significance level $\alpha$, an exact two-sided binomial test determines which matching counts are accepted under the null. The corresponding bit accuracies form the discrete acceptance set $\mathcal{A}_{N,\alpha}$. Here, acceptance means that the test does not detect the watermark; it does not establish that all watermark information has been erased. The exact decision rule used in this work is specified in \autoref{sec:threatmodel}.

\section{Related Work}\label{sec:related}

\subsection{Watermarking for AI-Generated Content}\label{sec:rw_wm}
Classical watermarking embeds signals through spread-spectrum or quantization-based constructions~\cite{cox1997,chen2001}. Learned watermarking changes the embedding mechanism, but retains the need to trade distortion against reliable detection. Watermarking schemes for generative models fall into two families according to where the mark is injected. \emph{Endogenous} schemes couple the watermark to the generation process itself. Stable Signature fine-tunes the VAE decoder of a latent diffusion model so that every image it produces carries a fixed multi-bit signature~\cite{fernandez2023stable}; GAN fingerprinting embeds a scalable identifier into the generator's parameters so that outputs remain attributable to a specific model instance~\cite{yu2021artificial}; and initial-noise schemes such as Tree-Ring and Gaussian Shading plant a structured pattern in the sampling seed and recover it by DDIM inversion~\cite{song2021ddim}~\cite{wen2023treering,yang2024gaussian}. \emph{Exogenous} schemes instead treat watermarking as a post-hoc image-processing step, training an encoder--decoder pair to hide and recover a bit string under a differentiable noise layer, as in HiDDeN and its noise-robust successor MBRS~\cite{zhu2018hidden,jia2021mbrs}. The two families differ in where the watermark is embedded
and how much of the pipeline the provider controls. The four
systems evaluated in this work expose a common soft per-bit
decoding interface, with agreement against an $N$-bit key
measured by \autoref{eq:bitacc}. Our attack therefore does not target a particular embedding mechanism, and we evaluate it on fragile and robust endogenous schemes and on post-hoc ones; broader taxonomies of the design space are available in recent surveys~\cite{zhao2025sok}.

Beyond these evaluated systems, StegaStamp targets physical capture robustness, while SteganoGAN studies high-capacity learned hiding~\cite{tancik2020,zhang2019stegano}. RoSteALS uses frozen autoencoder latents for embedding, and TrustMark supports robust watermarking at arbitrary resolution~\cite{bui2023,bui2025}. Watermark Anything extends decoding to localized messages, and VINE studies robustness against generative image editing~\cite{sander2025,lu2025vine}. These works broaden the embedding and threat-model landscape; they are related work, not additional victims evaluated here.

\subsection{Watermark Removal Attacks}\label{sec:rw_attack}

Watermark removal attacks seek to prevent detection while preserving image content. WEvade optimizes image perturbations using watermark-decoder feedback~\cite{jiang2023wevade}. Its WEvade-W-II variant addresses double-tail detection by targeting a randomly sampled watermark and searching over the perturbation budget. UnMarker disrupts spectral structures without detector feedback~\cite{kassis2025unmarker}, while regeneration attacks weaken watermarks through learned compression or diffusion-based reconstruction~\cite{zhao2024invisible,liu2025ctrlregen}. Lee et al.\ investigate how information near detection boundaries can support watermark removal~\cite{lee2025boundary}.

WEvade-W-II is the closest comparison because it explicitly accounts for detectable watermark inversion. LiBRA differs by constraining average decoding confidence rather than prescribing a target for every bit. Its symmetric objective corrects deviations on both sides of a near-chance target, while leaving individual bit configurations flexible so that fidelity constraints can favor less damaging changes. LiBRA jointly optimizes watermark and fidelity losses through bounded latent perturbations, without the outer perturbation-budget search used by WEvade-W-II.

Our matched latent-space comparisons show that this optimization procedure reduces runtime while maintaining similar evasion and image fidelity under the tested setting (\autoref{sec:exp_wevade}). Beyond this comparison, we evaluate both generation-integrated and post-hoc watermarking systems, quantify the gap between soft-confidence targets and actual detection outcomes, and examine adaptation to an additional detection statistic. The main LiBRA setting assumes knowledge or a reliable estimate of the watermark key, whereas WEvade-W-II does not require the true key.

\subsection{Adaptive Attacks and Evaluation Practice}\label{sec:rw_adaptive}
Research on adversarial robustness emphasizes evaluating defenses against attacks that account for their mechanisms~\cite{athalye2018,tramer2020,croce2020}. For watermark removal, this requires assessing success against the provider's actual detection rule. Low bit accuracy alone is insufficient when inverted watermarks remain detectable.

We evaluate image-level detection outcomes and distinguish them from the soft confidence optimized by the attack. In \autoref{sec:exp_adaptive}, we further examine a detector that checks confidence dispersion and an adaptive attack that accounts for this additional statistic.

\section{Proposed Method}\label{sec:threat}
\raggedbottom
\label{sec:method}
\subsection{Threat Model}\label{sec:threatmodel}
We consider a watermarking pipeline in which a provider embeds a secret watermark into an image and later uses a dedicated detector to verify its presence. The adversary seeks to modify the watermarked image so that it evades detection while preserving image quality. We specify the provider's detection rule and the adversary's capabilities below.

\subsubsection{The Victim's Goal}
The victim is the model provider, who wants every image emitted by its service to remain attributable while staying visually unchanged. Embedding is performed with a secret key $w\in\{-1,+1\}^{N}$ that only the provider's detector $D_m$ can read. Given an image $x'$, the detector produces a soft output $d=D_m(x')\in\mathbb{R}^{N}$, from which the per-image bit accuracy follows as
\begin{equation}
a(x')=\frac{1}{N}\sum_{i=1}^{N}\mathbf{1}\!\left[\operatorname{sign}(d_i)=w_i\right].
\label{eq:bitacc}
\end{equation}
How the provider maps $a(x')$ to a decision is central to our design. A naive \emph{one-sided} detector declares an image watermarked when $a(x')>0.5$, so an attack that merely drives $a$ below $0.5$ already counts as successful. A security-conscious provider instead applies an \emph{exact two-sided binomial test}. Under the null hypothesis that $x'$ is unwatermarked, the number of matching bits $K=N\,a(x')$ follows $\mathrm{Binomial}(N,0.5)$. Let $p_{\mathrm{exact}}(k;N)$ denote the exact two-sided binomial $p$-value for observing $k$ matches. The image is flagged when $p_{\mathrm{exact}}(k;N)<\alpha$ and is accepted as unwatermarked only when
\begin{equation}
a(x')\in\mathcal{A}_{N,\alpha}
=\left\{\frac{k}{N}\;\middle|\;
\begin{gathered}
k\in\{0,\ldots,N\},\\
p_{\mathrm{exact}}(k;N)\ge\alpha
\end{gathered}
\right\}.
\label{eq:band}
\end{equation}
Because the tested statistic is discrete, $\mathcal{A}_{N,\alpha}$ is a finite set of attainable bit accuracies rather than a continuous normal-approximation interval. Under the symmetric null considered here, the accepted integer counts $k$ are contiguous, so we report the smallest and largest accepted counts as its boundaries. The provider detects the watermark whenever $a(x')$ falls outside this set, since a statistically significant deviation from chance in either direction indicates an association with the key. A fully inverted watermark ($a\approx0$) therefore remains detectable, just as an unchanged watermark ($a\approx1$) does, because inversion preserves the key information in negated form. The provider also requires watermark embedding to preserve image quality and keeps both the key and the watermarked decoder confidential.

\subsubsection{The Adversary's Goal}
The adversary is a legitimate or semi-legitimate service user, such as a platform user, partner, or insider with limited privileges. The adversary obtains watermarked images through normal inference and seeks to remove their watermarks so that the content can be redistributed or misused without being traced. Given a watermarked image $x_0$, the adversary seeks a perceptually close image $x'$ that simultaneously achieves \emph{effectiveness}, meaning the detector cannot assert that the watermark is present, and \emph{fidelity}, meaning the image remains visually indistinguishable from $x_0$ under SSIM and LPIPS. Because the provider is assumed to run the two-sided test \autoref{eq:band}, effectiveness is \emph{not} the same as flipping the watermark: the adversary must land $a(x')$ \emph{inside} the exact acceptance set $\mathcal{A}_{N,\alpha}$ rather than push it to zero. Writing $L_w=E(x_0)$ for the latent of the watermarked image and $\delta$ for a perturbation confined to a frequency-guided spatial mask $M^{\mathrm{s}}$ (\autoref{sec:mask}) and kept within the admissible latent range, the decoded image is$x'=\operatorname{clip}\big(D(L_w+M^{\mathrm{s}}\odot\delta),-1,1\big)$ and the adversary solves
\begin{equation}
\min_{\delta}\;\lVert M^{\mathrm{s}}\odot\delta\rVert_2^2
\quad\text{s.t.}\quad
a(x')\in\mathcal{A}_{N,\alpha}.
\label{eq:threat_opt}
\end{equation}
In \autoref{eq:threat_opt}, the adversary seeks a small latent perturbation that places the attacked image within the detector's exact acceptance set defined in \autoref{eq:band}. The differentiable formulation and optimization procedure are presented in \autoref{sec:overview}--\autoref{sec:overall}.

\subsubsection{The Adversary's Capabilities}
We assume a white-box but resource-limited adversary. It (i) can query the detector $D_m$ and back-propagate through it, so gradients of the detection confidence with respect to the image are available; (ii) owns a publicly available standard VAE with encoder $E$ and decoder $D$, but \emph{not} the provider's secret watermarked decoder, so it must encode and decode through the clean VAE and can never re-embed or re-synthesize the watermark; (iii) can access or reliably estimate the key $w$, which specifies the key-aware setting evaluated in this work; and (iv) operates under a limited compute and memory budget, which motivates optimizing in the compact latent space to reduce the number of optimization variables. Conversely, the adversary does not interfere with the generation process, does not hold the provider's secret watermarking components, cannot modify or retrain the detector, and cannot observe the provider's decision threshold $\alpha$ directly---it can infer the exact acceptance set from $N$ and the assumed significance level via \autoref{eq:band}.

\autoref{tab:threat} summarizes these capabilities.

\begin{table}[t!]
\centering
\caption{Adversary capabilities assumed in this work.}
\label{tab:threat}
\renewcommand{\arraystretch}{1.1}
\begin{tabular}{lc}
\toprule
Capability & Assumed \\
\midrule
Query the detector $D_m$ on chosen images & yes \\
\rowcolor{gray!10}
Back-propagate gradients through $D_m$ & yes \\
Know or estimate the key $w$ & yes \\
\rowcolor{gray!10}
Know the key length $N$ & yes \\
Use a public, non-watermarked VAE $(E,D)$ & yes \\
\midrule
\rowcolor{gray!10}
Access the provider's watermarked decoder & no \\
Modify or retrain the detector & no \\
\rowcolor{gray!10}
Interfere with the generation process & no \\
Observe the provider's significance level $\alpha$ & no \\
\rowcolor{gray!10}
Re-embed or re-synthesize a watermark & no \\
\bottomrule
\end{tabular}
\end{table}

\subsection{Overview}\label{sec:overview}
\begin{figure*}[!t]
\centering
\includegraphics[width=\textwidth]{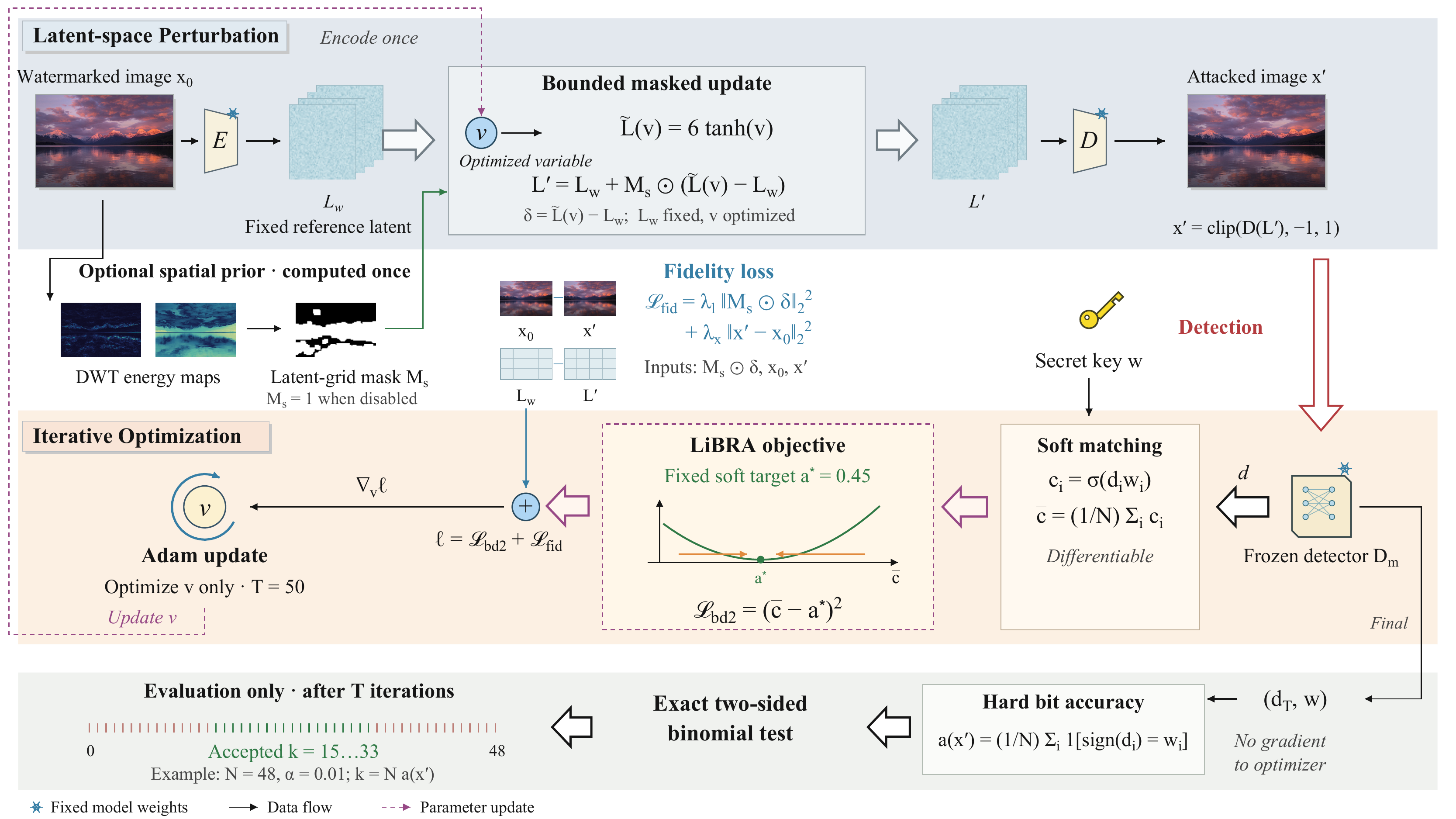}
\caption{Overview of the attack. Stage~1 builds an optional spatial prior $M^{\mathrm{s}}$ from the image's DWT energy; Stage~2 perturbs the latent code of a public autoencoder inside that prior and decodes the attacked image $x'$; Stage~3 queries the detector and scores the decoded bits with the watermark objective. Dashed arrows denote parameter updates. The symmetric objective guides mean confidence toward a near-chance target and counteracts excessive inversion, while fidelity penalties constrain changes to the latent representation and reconstructed image.}
\label{fig:pipeline}
\end{figure*}

\autoref{fig:pipeline} summarizes LiBRA. The method combines bounded latent optimization with a symmetric mean-confidence objective. Latent and image-level fidelity penalties constrain changes, while an optional spatial prior restricts the modified regions. The watermark objective guides mean confidence toward a near-chance target and counteracts excessive inversion. To isolate its contribution, the main objective comparison uses a common latent platform. Separate experiments examine the effects of optimization space, spatial prior, key knowledge, and other supporting components.

\subsection{Attack Platform}\label{sec:box}
\subsubsection{Bounded latent perturbation}
The attacker encodes the watermarked image with a clean public autoencoder, $L_w=E(x_0)$, and optimizes a perturbation of that code. We enforce the latent bounds $[L_{\min},L_{\max}]$ through a $\tanh$ reparameterization, allowing unconstrained optimization without a separate projection step~\cite{carlini2017towards}:
\begin{equation}
\tilde{L}(v)=\frac{L_{\max}-L_{\min}}{2}\big(\tanh(v)+1\big)+L_{\min},
\label{eq:box}
\end{equation}
and optimize the free variable $v$. It is initialized so that $\tilde{L}(v_0)=L_w$, i.e.\ $v_0=\operatorname{artanh}\!\big(2(L_w-L_{\min})/(L_{\max}-L_{\min})-1\big)$, so the attack starts from the unperturbed code rather than from zero.

\subsubsection{Fidelity}
The perturbation $\delta=\tilde{L}(v)-L_w$ is penalized both in the latent code and in the decoded image $x'=\operatorname{clip}\big(D(L_w+M^{\mathrm{s}}\odot\delta),-1,1\big)$,
\begin{equation}
L_{\mathrm{fid}}=\lambda_{l}\,\lVert M^{\mathrm{s}}\odot\delta\rVert_2^2+\lambda_{x}\,\lVert x'-x_0\rVert_2^2 .
\label{eq:freq}
\end{equation}
Our implementation computes the image term as a sum of squared errors over the four subbands of a one-level orthonormal Haar decomposition with equal weights. Because that transform is orthonormal the two quantities are identical, so we state the term directly in the pixel domain rather than presenting it as a separate frequency-domain constraint.

At each iteration, we clip the decoded image to $[-1,1]$
before watermark decoding and image-fidelity evaluation.
The final output uses the same clipping operation.

\subsection{Heuristic Spatial Prior}\label{sec:mask}
\begin{figure}[t!]
\centering
\includegraphics[width=\linewidth]{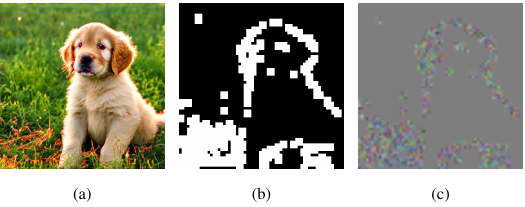}
\caption{The heuristic spatial prior: (a) watermarked input; (b) spatial mask; (c) perturbation visualization. The mask describes image content, not watermark localization.}
\label{fig:mask}
\end{figure}

The platform can restrict the perturbation to part of the latent grid. The mask is a heuristic prior on where a perturbation is least visible---textured and structured regions---and not a localization of watermark energy: it is computed from image content alone, and nothing in its construction refers to the watermark. We apply a one-level Haar DWT to $x_0$ and form high- and low-frequency energy maps
\begin{equation}
E_H=\sqrt{Y_{LH}^{2}+Y_{HL}^{2}+Y_{HH}^{2}},\qquad E_L=\lvert Y_L\rvert,
\label{eq:energy}
\end{equation}
threshold them at fractions $\tau_h,\tau_l$ of their maxima, take the union, fill holes with a $k\times k$ morphological closing, resize the result to the latent grid and dilate it by max-pooling over a $(2p{+}1)\times(2p{+}1)$ neighbourhood,
\begin{equation}
M^{\mathrm{s}}(i,j)=\max_{(m,n)\in\mathcal{N}_p(i,j)} M^{\downarrow}(m,n).
\label{eq:expand}
\end{equation}
\autoref{fig:mask} shows an example. Two caveats follow from the construction. The low-frequency branch thresholds the magnitude of the approximation band, which is a brightness measure, so on a flat bright image it selects everything. And the mask reduces the number of \emph{free} coordinates, not the per-iteration cost, because the full latent is still decoded and the detector still runs on the full image. Whether the prior is worth its parameters is therefore an empirical question, which \autoref{sec:exp_abl} answers by comparing it with no mask, a random mask of equal area and a plain edge mask.

\subsection{Watermark Objective}\label{sec:wmloss}
\begin{figure}[t!]
\centering
\includegraphics[width=.9\linewidth]{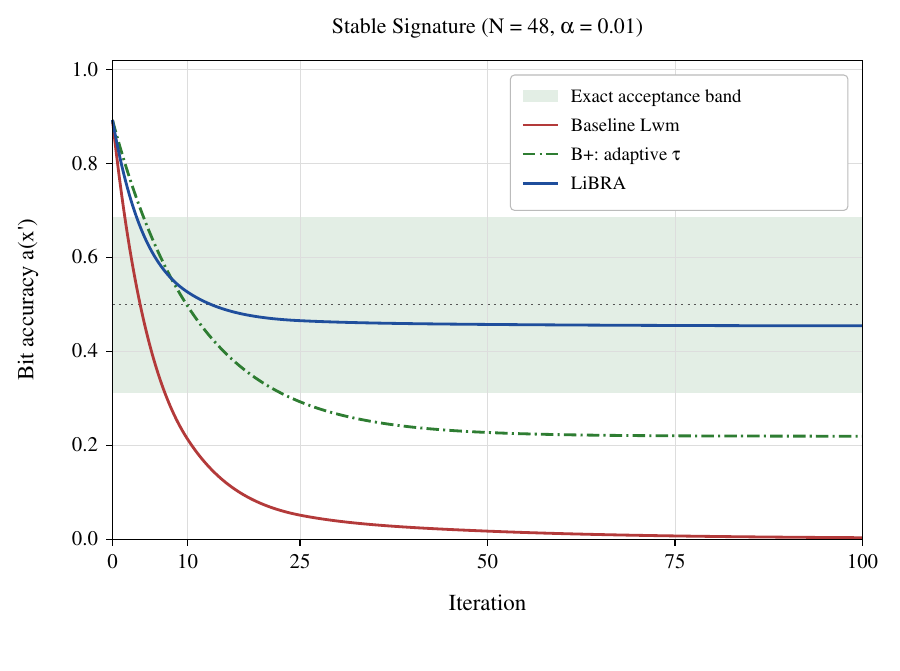}
\caption{Bit-accuracy trajectories under the baseline, the adaptive per-bit variant (B+) and the two-sided objective (LiBRA).}
\label{fig:boundary}
\end{figure}

Let $d=D_m(x')$, $s_i=d_iw_i$ the per-bit margin and $c_i=\sigma(s_i)$ the per-bit confidence, so that $c_i>0.5$ exactly when bit $i$ is decoded correctly. The detector tests the hard bit accuracy $a(x')$ of \autoref{eq:bitacc}, which is piecewise constant; every objective below therefore acts on the soft confidences instead. \autoref{fig:boundary} shows the trajectories they produce.

\subsubsection{Baseline}
The standard objective minimizes the mean confidence,
\begin{equation}
L_{\mathrm{wm}}=\frac{1}{N}\sum_{i=1}^{N}\sigma(d_i w_i).
\label{eq:base_wm}
\end{equation}
It rewards lowering every margin without limit, so its natural end state is a fully inverted decoding, $a(x')\to0$, which lies outside the acceptance set \autoref{eq:band} and remains detectable by a two-sided test.

\subsubsection{Two-sided objective}
We instead penalize the distance of the mean confidence from a target $a^{\star}$ chosen between the smallest and largest accepted hard bit accuracies,
\begin{equation}
L_{\mathrm{bd2}}=\Big(\frac{1}{N}\sum_{i=1}^{N}\sigma(d_i w_i)-a^{\star}\Big)^{2}.
\label{eq:bpp}
\end{equation}
The loss is symmetric: it pushes the confidences down while their mean is above $a^{\star}$ and back up when it falls below, so an overshoot is corrected rather than left in place. This objective forms the core of LiBRA. It guides mean confidence toward a near-chance target and penalizes deviations in both directions, counteracting excessive inversion while leaving individual bit configurations flexible for fidelity preservation.

\subsubsection{Per-bit alternatives}
Two simpler modifications of the baseline suggest themselves, and we evaluate both to establish that the aggregate form of \autoref{eq:bpp} is needed. The first (B) stops pushing a bit once its confidence falls below a threshold $\tau$,
\begin{equation}
L_{\mathrm{bd}}=\frac{1}{N}\sum_{i=1}^{N}\operatorname{ReLU}\!\big(\sigma(d_i w_i)-\tau\big),
\label{eq:b}
\end{equation}
and the second (B+) adapts $\tau$ online from the running bit accuracy $a_t$,
\begin{equation}
\tau_{t+1}=\operatorname{clip}\!\big(\tau_t-\eta\,(a_t-a^{\star}),\;\tau_{\min},\;\tau_{\max}\big).
\label{eq:bplus}
\end{equation}
Both constrain bits individually, and that is where they fail. For any $\tau\le0.5$ the zero set of \autoref{eq:b} consists of decodings in which \emph{every} bit is wrong, so the hinge still drives toward full inversion and only slows down on the way; the controller can raise $\tau$ to slow the descent further, but the per-bit loss provides no direct incentive to restore a bit once its confidence falls below the threshold. B and B+ are therefore ablations, not components of the proposed method.

\subsection{What the Objectives Control}\label{sec:analysis}
The objectives of \autoref{sec:wmloss} are easiest to compare through their gradients with respect to the per-bit margins $s_i=d_iw_i$. Two cautions apply to everything in this subsection. First, the optimization variable is the shared latent code $v$, not the margins: with $J=\partial s/\partial v$, a small step on the watermark term alone changes the margins by $\Delta s\approx-\gamma JJ^{\top}\nabla_{s}L$, so a positive partial derivative $\partial L/\partial s_i$ does not guarantee that $s_i$ itself decreases, and the fidelity term couples the coordinates further. The statements below describe what each objective \emph{rewards}, not how the optimizer converges. Second, all three act on soft confidences, while the detector tests the hard bit accuracy.

\begin{proposition}[Baseline]\label{prop:base}
For $L_{\mathrm{wm}}$ of \autoref{eq:base_wm}, $\partial L_{\mathrm{wm}}/\partial s_i=\tfrac1N\sigma(s_i)\big(1-\sigma(s_i)\big)>0$ for every finite $s_i$. Hence $L_{\mathrm{wm}}$ has no stationary point at finite margins, and its infimum is approached only as every $s_i\to-\infty$, i.e.\ at $a(x')=0$.
\end{proposition}

\begin{proposition}[Per-bit hinge]\label{prop:hinge}
For $L_{\mathrm{bd}}$ of \autoref{eq:b} with $\tau\le0.5$, the loss is zero exactly on $\mathcal{S}_\tau=\{s:\,c_i\le\tau\ \forall i\}$. Every point of $\mathcal{S}_\tau$ has $c_i\le0.5$ for all $i$, and therefore $a(x')=0$.
\end{proposition}

\noindent \autoref{prop:hinge} is the precise sense in which the per-bit variants do not solve the problem. The hinge stops pushing a bit once it has crossed $\tau$, which slows the descent, but its minimizers are still fully inverted decodings: a per-bit decision boundary has been used in place of the image-level acceptance set, and the two define different sets. Adapting $\tau$ through \autoref{eq:bplus} moves where the hinge stops, not what its minimizers are; and because $a_t$ takes only the values $k/N$, the controller need not reach $a_t=a^{\star}$ exactly.

\begin{proposition}[Two-sided objective]\label{prop:twosided}
For $L_{\mathrm{bd2}}$ of \autoref{eq:bpp}, $\partial L_{\mathrm{bd2}}/\partial s_i=\tfrac{2}{N}(\bar c-a^{\star})\,\sigma'(s_i)$ with $\bar c=\frac1N\sum_i c_i$. Every partial derivative has the sign of $\bar c-a^{\star}$, and the loss is zero exactly on $\{s:\bar c=a^{\star}\}$.
\end{proposition}

\noindent The symmetry is what distinguishes LiBRA from the hinge: an overshoot reverses the sign of the gradient instead of zeroing it. What the proposition does \emph{not} give is evasion. The zero set is a level set of the mean confidence, but hard
bit accuracy is not determined by this mean. For example,
80 bits with confidence 0.55 and 20 bits with confidence 0.05
yield a mean confidence of 0.45 but a hard bit accuracy of 0.80. Nothing in the loss rules such points out. What can make them rare in practice is that the margins are coupled through the shared latent and tend to move together, which is a property of the victim detectors rather than of the objective. We therefore do not claim that reaching $\bar c=a^{\star}$ places an image inside the acceptance set, nor that attacked images are statistically indistinguishable from unwatermarked ones, which a single test failing to reject would not establish in any case. \autoref{sec:exp_gap} measures how often the surrogate and the tested statistic disagree on the victim systems.

\subsection{Overall Objective and Algorithm}\label{sec:overall}
The attack solves
\begin{equation}
\min_{v}\;\; L_{\mathrm{bd2}}\big(D_m(x')\big) + L_{\mathrm{fid}},
\label{eq:total}
\end{equation}
with $x'$ and $L_{\mathrm{fid}}$ as in \autoref{eq:freq}; the baseline and the per-bit variants replace $L_{\mathrm{bd2}}$ by \autoref{eq:base_wm} or \autoref{eq:b}. \autoref{alg:attack} gives the procedure, using Adam~\cite{kingma2015adam}. Each iteration costs one decoder and one detector forward--backward pass, $\mathcal{O}\!\big(T(C_D+C_{D_m})\big)$ in total. Optimizing the latent reduces the number of free variables by $48\times$ at $512\times512$ relative to the pixel grid, but it does not by itself reduce the per-iteration cost, which the decoder dominates; measured cost is reported in \autoref{sec:setup}.

\begin{algorithm}[t!]
\caption{LiBRA: Latent In-band Bidirectional Removal Attack}
\label{alg:attack}
\begin{algorithmic}[1]
\Require watermarked image $x_0$, key $w$, detector $D_m$, VAE $(E,D)$, iterations $T$, target $a^{\star}$
\Ensure attacked image $x'$
\State $L_w \gets E(x_0)$ 
\State build $M^{\mathrm{s}}$ by \autoref{eq:energy}--\autoref{eq:expand}, or set $M^{\mathrm{s}}=\mathbf{1}$ 
\State $v \gets \operatorname{artanh}\!\big(2(L_w-L_{\min})/(L_{\max}-L_{\min})-1\big)$ 
\For{$t = 1$ \textbf{to} $T$}
  \State $\delta \gets \tilde{L}(v)-L_w$
\State $x' \gets \operatorname{clip}\big(D(L_w+M^{\mathrm{s}}\odot\delta),-1,1\big)$
  \State $d \gets D_m(x')$
  \State $\ell \gets L_{\mathrm{bd2}}(d) + L_{\mathrm{fid}}(\delta,x')$ 
  \State $v \gets \textsc{AdamUpdate}(v, \nabla_v \ell)$
\EndFor
\State $L' \gets L_w+M^{\mathrm{s}}\odot(\tilde{L}(v)-L_w)$
\State \Return $\operatorname{clip}\big(D(L'),-1,1\big)$
\end{algorithmic}
\end{algorithm}

\section{Experiments}\label{sec:exp}

\setlength{\textfloatsep}{12pt plus 2pt minus 2pt}
\setlength{\intextsep}{8pt plus 2pt minus 2pt}
\setlength{\dbltextfloatsep}{12pt plus 2pt minus 2pt}

\subsection{Experimental Setup}\label{sec:setup}

\subsubsection{Watermarking systems and datasets}
We evaluate the released checkpoints of Stable Signature, YU1, HiDDeN, and MBRS~\cite{fernandez2023stable,yu2021artificial,zhu2018hidden,jia2021mbrs}, using $5{,}000$ watermarked images per system. Stable Signature images are generated at $512\times512$ from MS-COCO~2017 validation captions~\cite{lin2014coco}. YU1 uses fingerprinted CelebA test images~\cite{liu2015celeba} at $128\times128$. HiDDeN and MBRS watermark the same $5{,}000$ MS-COCO validation images at $256\times256$. Each configuration in the four-system main experiment therefore covers $20{,}000$ images, with results averaged separately for each system.

\subsubsection{Implementation}
The attacker uses the public, non-watermarked Stable Diffusion~2.1-base VAE without model training. Experiments run on Ubuntu~22.04 with eight NVIDIA A100 GPUs (80\,GB each), using PyTorch~2.4.1, CUDA~12.4, PyWavelets, and AlexNet-based LPIPS. Each image is processed on one GPU; multiple GPUs process different images in parallel. Reported per-image runtimes are 12.4\,s at $512\times512$ and 1.3\,s at $128\times128$. Resolution-specific runtime and memory measurements appear in \autoref{tab:res}.

\subsubsection{Attack configuration}
\autoref{tab:hp} lists the default hyperparameters. Unless otherwise stated, objective variants share the optimizer, spatial restriction, and fidelity terms; only the reproduced baseline includes the sign-consistency term. We use $a^{\star}=0.45$, a fixed random seed, and no per-image tuning. Every configuration includes all $5{,}000$ images for its system.

\begin{table}[t!]
\centering
\footnotesize
\setlength{\tabcolsep}{4pt}
\caption{Attack hyperparameters.}
\label{tab:hp}
\begin{tabular}{llc}
\toprule
Symbol & Meaning & Value \\
\midrule
$T$ & optimization iterations & $50$ \\
\rowcolor{gray!10}
--- & optimizer / learning rate & Adam / $0.01$ \\
$\tau_h$, $\tau_l$ & mask thresholds (high, low freq.) & $0.3$, $0.5$ \\
\rowcolor{gray!10}
$k$, $p$ & closing kernel side length, expansion radius & $5$, $2$ \\
$\lambda_{l}$ & weight of $\lVert M^{\mathrm{s}}\odot\delta\rVert_2^2$ & $10^{-2}$ \\
\rowcolor{gray!10}
$\lambda_{x}$ & weight of $\lVert x'-x_0\rVert_2^2$ & $10^{-4}$ \\
$\lambda_{\mathrm{bit}}$ & sign-consistency (baseline / others) & $1$ / $0$ \\
\rowcolor{gray!10}
$\tau_0$ & initial hinge threshold (B, B+) & $0.5$ \\
$\eta$ & threshold update gain (B+) & $0.5$ \\
\rowcolor{gray!10}
$a^{\star}$ & target mean confidence & $0.45$ \\
$\alpha$ & detector significance level & $0.01$ \\
\rowcolor{gray!10}
$[L_{\min},L_{\max}]$ & admissible latent range & $[-6,6]$ \\
\bottomrule
\end{tabular}
\end{table}

\subsubsection{Detectors and metrics}
We apply the naive one-sided rule $a(x')>0.5$ and the exact two-sided binomial test at $\alpha=0.01$ defined in \autoref{sec:threatmodel}. Accepted bit counts are listed in \autoref{tab:main}. We denote the proportions of images flagged by these rules as $\mathrm{DR}_1$ and $\mathrm{DR}_2$, respectively, and report both as fractions between $0$ and $1$. Lower detection rates indicate better evasion. Bit ACC is mean hard bit accuracy and has no monotonic preference. Fidelity is measured against the watermarked input using PSNR in dB, SSIM~\cite{wang2004ssim}, and LPIPS~\cite{zhang2018lpips}. Higher PSNR and SSIM and lower LPIPS indicate better fidelity.

\subsection{Main Results}\label{sec:exp_main}
\autoref{tab:main} compares the four objectives on four watermarking systems. The first column lists the watermark length and the exact accepted bit-count range for each system. Arrows indicate the preferred direction of each metric. Bold metric values mark the best detection and fidelity results among the attacked configurations for each system. Every attacked row has zero $\mathrm{DR}_1$ under the naive one-sided rule. The two-sided test separates these apparently identical outcomes: the inversion baseline is detected on every image, whereas LiBRA reaches detection rates between $0$ and $0.05$.

\begin{table*}[!t]
\centering
\footnotesize
\setlength{\tabcolsep}{5pt}
\caption{Main results across four watermarking systems.}
\label{tab:main}
\begin{tabular}{llcccccc}
\toprule
Scheme ($N$, accepted $k$)
& Configuration
& $\mathrm{DR}_1\downarrow$
& Bit ACC
& PSNR (dB)$\uparrow$
& SSIM$\uparrow$
& LPIPS$\downarrow$
& $\mathrm{DR}_2\downarrow$ \\
\midrule
\multirow{5}{*}{\shortstack[l]{Stable Signature\\
$N=48$, $k=15{:}33$}}
 & No attack
 & 1.00 & 0.996 & --- & 1.000 & 0.000 & 1.00 \\
 & Baseline
 & 0.00 & 0.018 & 35.1 & 0.942 & 0.074 & 1.00 \\
 & B (fixed $\tau$)
 & 0.00 & 0.091 & 36.0 & 0.951 & 0.066 & 1.00 \\
 & B+ (adaptive $\tau$)
 & 0.00 & 0.228 & 36.8 & 0.958 & 0.060 & 1.00 \\
 & \textbf{LiBRA (two-sided)}
 & 0.00 & 0.458 & \textbf{38.0}
 & \textbf{0.970} & \textbf{0.046} & \textbf{0.03} \\
\midrule
\multirow{5}{*}{\shortstack[l]{YU1\\
$N=100$, $k=37{:}63$}}
 & No attack
 & 1.00 & 0.998 & --- & 1.000 & 0.000 & 1.00 \\
 & Baseline
 & 0.00 & 0.001 & 26.9 & 0.740 & 0.095 & 1.00 \\
 & B (fixed $\tau$)
 & 0.00 & 0.061 & 27.8 & 0.758 & 0.084 & 1.00 \\
 & B+ (adaptive $\tau$)
 & 0.00 & 0.229 & 28.6 & 0.776 & 0.069 & 0.98 \\
 & \textbf{LiBRA (two-sided)}
 & 0.00 & 0.433 & \textbf{29.9}
 & \textbf{0.982} & \textbf{0.031} & \textbf{0.00} \\
\midrule
\multirow{5}{*}{\shortstack[l]{HiDDeN\\
$N=30$, $k=8{:}22$}}
 & No attack
 & 1.00 & 0.990 & --- & 1.000 & 0.000 & 1.00 \\
 & Baseline
 & 0.00 & 0.003 & 34.8 & 0.928 & 0.082 & 1.00 \\
 & B (fixed $\tau$)
 & 0.00 & 0.067 & 35.6 & 0.938 & 0.073 & 0.98 \\
 & B+ (adaptive $\tau$)
 & 0.00 & 0.200 & 36.4 & 0.947 & \textbf{0.065} & 0.82 \\
 & \textbf{LiBRA (two-sided)}
 & 0.00 & 0.433 & \textbf{37.6}
 & \textbf{0.972} & \textbf{0.065} & \textbf{0.04} \\
\midrule
\multirow{5}{*}{\shortstack[l]{MBRS\\
$N=64$, $k=22{:}42$}}
 & No attack
 & 1.00 & 0.995 & --- & 1.000 & 0.000 & 1.00 \\
 & Baseline
 & 0.00 & 0.006 & 35.0 & 0.934 & 0.079 & 1.00 \\
 & B (fixed $\tau$)
 & 0.00 & 0.083 & 35.8 & 0.943 & 0.070 & 1.00 \\
 & B+ (adaptive $\tau$)
 & 0.00 & 0.241 & 36.5 & 0.951 & \textbf{0.063} & 1.00 \\
 & \textbf{LiBRA (two-sided)}
 & 0.00 & 0.445 & \textbf{37.3}
 & \textbf{0.971} & \textbf{0.063} & \textbf{0.05} \\
\bottomrule
\end{tabular}
\end{table*}

On YU1, the baseline reaches a mean Bit ACC of $0.001$
and a two-sided detection rate of $1.00$. B+ increases
mean Bit ACC to $0.229$, but $98\%$ of images remain
detectable. LiBRA reaches a mean Bit ACC of $0.433$
with no observed two-sided detections. It also increases
SSIM from $0.740$ to $0.982$ and reduces LPIPS from
$0.095$ to $0.031$ relative to the baseline.

On Stable Signature, LiBRA reduces the two-sided detection
rate from $1.00$ to $0.03$, while increasing SSIM from
$0.942$ to $0.970$ and reducing LPIPS from $0.074$ to
$0.046$. On HiDDeN and MBRS, LiBRA reaches mean bit
accuracies of $0.433$ and $0.445$, with two-sided detection
rates of $0.04$ and $0.05$, respectively. The corresponding
SSIM values are $0.972$ and $0.971$, and LPIPS values are
$0.065$ and $0.063$. Both systems show higher SSIM and
lower LPIPS than the inversion baseline.

Means must nevertheless be read with the detection-rate column. For HiDDeN, the exact accepted range is $k=8{:}22$ at $N=30$. B+ has mean Bit ACC $0.200$, outside that range, although $18\%$ of images are accepted; B similarly leaves $2\%$ accepted despite a mean of $0.067$. \autoref{sec:exp_dist} explains why the per-image count distribution is needed.

\subsection{Distribution of the Achieved Operating Point}\label{sec:exp_dist}
\begin{figure}[t!]
\centering
\includegraphics[width=0.94\columnwidth]{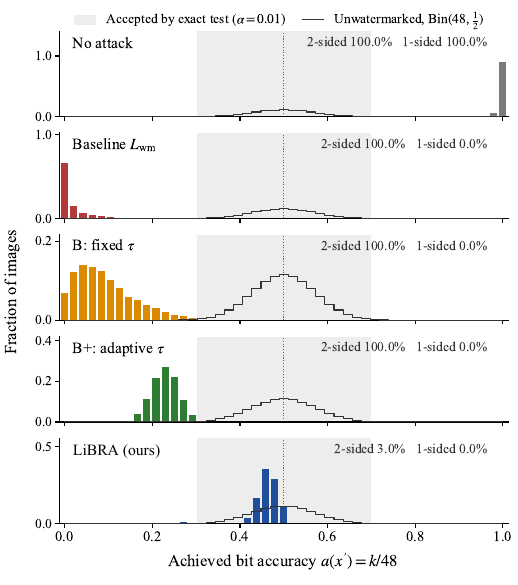}
\caption{Distribution of achieved bit accuracy on Stable Signature over 5,000 images, on the 48-bit grid. Panels compare no attack, baseline, B, B+, and LiBRA. Shading denotes exact accepted counts $k=15{:}33$ at $\alpha=0.01$; the black curve is the theoretical binomial null. Percentages report the two-sided and naive one-sided detection rates.}
\label{fig:hist}
\end{figure}

Mean bit accuracy alone does not determine the fraction of images detected: a mean inside the acceptance set is compatible with individual images outside it. For $N=48$, the statistic has only 49 attainable values, $k/48$. An empirical evaluation should therefore plot the count distribution and sum the mass on rejected counts. \autoref{fig:hist} shows the per-image distributions for the evaluated attack objectives. Its shaded region covers $k=15{:}33$, and its black curve is the theoretical binomial null.

\subsection{The Surrogate Gap}\label{sec:exp_gap}
\begin{figure}[t!]
    \centering
    \includegraphics[width=\columnwidth]
        {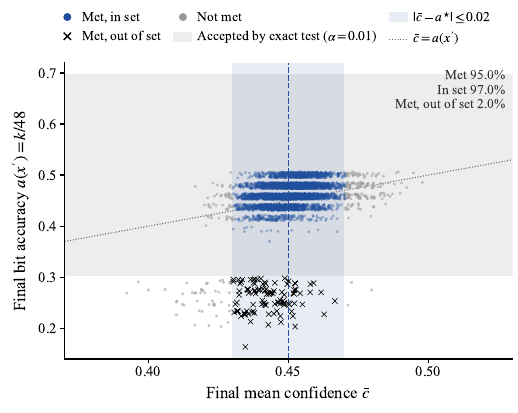}
    \caption{Relationship between final mean confidence and hard bit accuracy under LiBRA on Stable Signature ($N=48$). The vertical band marks $|\bar c-0.45|\le0.02$ and the horizontal band marks exact accepted bit counts. Crosses illustrate meeting the soft target while failing the hard test. Each point represents one attacked image, with horizontal and vertical jitter for visibility; categories and statistics use unjittered values.}
    \label{fig:gap}
\end{figure}

LiBRA optimizes mean soft confidence, while the detector evaluates hard bit accuracy. \autoref{fig:gap} plots these quantities for the attacked images, and \autoref{tab:gap} measures their agreement. ``Soft target met'' denotes $|\bar c-a^{\star}|\le0.02$, and ``Hard-test evasion'' denotes hard bit accuracy within the exact acceptance set. ``Target met, detected'' denotes images that meet the soft target but remain detectable. All three columns report fractions of the full evaluated image set.

\begin{table}[t!]
\centering
\footnotesize
\setlength{\tabcolsep}{4pt}
\caption{Soft-target and hard-test agreement.}
\label{tab:gap}
\begin{tabular}{lcccc}
\toprule
Scheme & $N$ & \shortstack{Soft target\\met} & \shortstack{Hard-test\\evasion} & \shortstack{Target met,\\detected} \\
\midrule
Stable Signature & 48 & 0.95 & 0.97 & 0.02 \\
\rowcolor{gray!10}
YU1 & 100 & 0.97 & 1.00 & 0.00 \\
HiDDeN & 30 & 0.93 & 0.96 & 0.03 \\
\rowcolor{gray!10}
MBRS & 64 & 0.94 & 0.95 & 0.03 \\
\bottomrule
\end{tabular}
\end{table}

Hard-test evasion ranges from $0.95$ to $1.00$ across the four systems. Images that meet the soft target but remain detectable account for $0.02$ of all images on Stable Signature, $0.00$ on YU1, and $0.03$ on both HiDDeN and MBRS. These results identify a specific difference between target attainment and successful removal: some images reach the desired mean confidence while retaining a detectable hard-bit pattern.

\subsection{Comparison with Existing Removal Attacks}\label{sec:exp_cmp}
\autoref{tab:cmp} places our objective against both families an adversary might otherwise reach for: the classical post-processing distortions that any deployed watermark is expected to survive, and the removal attacks discussed in \autoref{sec:rw_attack}.

\begin{table*}[!t]
\centering
\footnotesize
\setlength{\tabcolsep}{5pt}
\caption{Comparison with removal attacks and image transformations.}
\label{tab:cmp}
\begin{tabular}{llccccc}
\toprule
Family & Method & $\mathrm{DR}_1\downarrow$ & Bit ACC & SSIM$\uparrow$ & LPIPS$\downarrow$ & $\mathrm{DR}_2\downarrow$ \\
\midrule
\multirow{5}{*}{Post-processing}
 & Center crop ($10\%$)      & 1.00 & 0.943 & 0.618 & 0.312 & 1.00 \\
 & JPEG ($q=90$)             & 1.00 & 0.991 & 0.967 & 0.029 & 1.00 \\
 & Gaussian blur ($k=5$)     & 1.00 & 0.978 & 0.884 & 0.148 & 1.00 \\
 & Bilateral filter          & 1.00 & 0.969 & 0.912 & 0.106 & 1.00 \\
 & Quantization ($8$ levels) & 1.00 & 0.927 & 0.741 & 0.223 & 1.00 \\
\midrule
\multirow{6}{*}{Removal attacks}
 & WEvade-W-I (flip)~\cite{jiang2023wevade}     & 0.00 & 0.012 & 0.944 & 0.067 & 1.00 \\
 & WEvade-W-II ($a\!\approx\!0.5$)~\cite{jiang2023wevade} & 0.42 & 0.497 & 0.951 & 0.061 & 0.02 \\
 & UnMarker~\cite{kassis2025unmarker}          & 0.75 & 0.566 & 0.832 & 0.187 & 0.13 \\
 & VAE regeneration~\cite{zhao2024invisible,liu2025ctrlregen}     & 0.91 & 0.662 & 0.861 & 0.121 & 0.41 \\
 & Diffusion regeneration~\cite{zhao2024invisible,liu2025ctrlregen} & 0.68 & 0.548 & 0.793 & 0.169 & 0.09 \\
 & Latent attack, $L_{\mathrm{wm}}$ objective    & 0.00 & 0.018 & 0.942 & 0.074 & 1.00 \\
\midrule
Ours & \textbf{LiBRA (two-sided)}          & 0.00 & 0.458 & 0.970 & 0.046 & 0.03 \\
\bottomrule
\end{tabular}
\end{table*}

\autoref{tab:cmp} compares removal attacks and image transformations on $5{,}000$ Stable Signature images. All tested post-processing transformations have a two-sided detection rate of $1.00$. WEvade-W-I and the latent inversion baseline achieve zero detection under the naive one-sided rule but remain fully detectable under the two-sided test. LiBRA reaches a two-sided detection rate of $0.03$, compared with $0.02$ for pixel-domain WEvade-W-II, while increasing SSIM from $0.951$ to $0.970$ and reducing LPIPS from $0.061$ to $0.046$. The runtime comparison in \autoref{tab:wevade} further shows a reduction from $24.8$ to $12.4$~s per image. LiBRA therefore achieves higher image fidelity and halves the measured runtime at comparable two-sided evasion.

\subsection{Head-to-Head with WEvade-W-II}\label{sec:exp_wevade}
\autoref{tab:wevade} compares optimization space, optimization procedure, and reference-bit choice on Stable Signature. WEvade-W-II~\cite{jiang2023wevade} optimizes toward a random target bit string and searches over the perturbation budget. LiBRA uses a single optimization procedure with a symmetric mean-confidence objective and fidelity penalties. We evaluate WEvade-W-II in both pixel and latent spaces, and LiBRA with either a random reference bit string or the true watermark key. In the ``Space / reference'' column, ``random'' denotes a uniformly sampled bit string and ``key'' denotes the true key. The two latent-space rows using random references compare the procedures under a common perturbation space and reference setting.

\begin{table}[t!]
\centering
\footnotesize
\setlength{\tabcolsep}{3pt}
\renewcommand{\arraystretch}{1.1}
\caption{Comparison with WEvade-W-II.}
\label{tab:wevade}
\resizebox{\columnwidth}{!}{%
\begin{tabular}{@{}llcccc@{}}
\toprule
Method
& \shortstack{Space /\\reference}
& $\mathrm{DR}_2\downarrow$
& SSIM$\uparrow$
& LPIPS$\downarrow$
& \shortstack{Time\\(s/image)} \\
\midrule
WEvade-W-II
& pixel / random & 0.02 & 0.951 & 0.061 & 24.8 \\

WEvade-W-II
& latent / random & 0.03 & 0.965 & 0.050 & 118.6 \\
LiBRA
& latent / random & 0.04 & 0.967 & 0.048 & 12.5 \\

\textbf{LiBRA}
& latent / key & 0.03 & 0.970 & 0.046 & 12.4 \\
\bottomrule
\end{tabular}%
}
\end{table}

With a random reference, LiBRA achieves a two-sided detection rate of $0.04$, SSIM of $0.967$, and LPIPS of $0.048$, compared with $0.03$, $0.970$, and $0.046$ when using the true key. Under the common latent-space and random-reference setting, WEvade-W-II requires $118.6$~s per image, while LiBRA requires $12.5$~s, giving an approximately $9.5\times$ speedup. LiBRA also increases SSIM from $0.965$ to $0.967$ and reduces LPIPS from $0.050$ to $0.048$, with detection rates of $0.04$ and $0.03$, respectively. This comparison demonstrates the computational benefit of LiBRA's single optimization procedure at comparable evasion.

\subsection{Is Each Component Necessary?}\label{sec:exp_abl}
The attack platform carries several components besides the objective, and \autoref{sec:method} argues that none of them should be assumed necessary. \autoref{tab:abl} evaluates the components on Stable Signature. Unless otherwise stated, the configurations use the LiBRA objective and the fidelity loss in \autoref{eq:freq}. The ``LiBRA (frequency mask)'' row is the full configuration; the $+L_{\mathrm{bit}}$ and $+L_{\mathrm{dfl}}$ rows add the corresponding term to this configuration. The rows answer specific questions rather than enumerate combinations. Clean reconstruction without optimization measures how much of the effect comes from passing the image through the autoencoder at all. Full-latent optimization removes the spatial prior. The random and edge masks are matched to the frequency-guided mask in area to assess whether its construction provides benefits beyond restricting the perturbation region. Adding the sign-consistency term tests whether a per-bit flipping pressure helps or fights the aggregate target. The early-stopped baseline keeps \autoref{eq:base_wm} but returns the first iterate whose bit accuracy enters the band, which separates the effect of the loss shape from the effect of simply not optimizing past the band.

\begin{table}[t!]
\centering
\footnotesize
\setlength{\tabcolsep}{2.2pt}
\renewcommand{\arraystretch}{1.1}
\caption{Ablation study of attack components.}
\label{tab:abl}
\resizebox{\columnwidth}{!}{%
\begin{tabular}{@{}lccccc@{}}
\toprule
Configuration
& Bit ACC
& SSIM$\uparrow$
& LPIPS$\downarrow$
& $\mathrm{DR}_2\downarrow$
& \shortstack{Time\\(s/image)} \\
\midrule
VAE reconstruction only
& 0.893 & 0.976 & 0.039 & 1.00 & 0.2 \\

Full latent, no mask
& 0.461 & 0.962 & 0.053 & 0.03 & 12.1 \\

Random mask
& 0.452 & 0.964 & 0.051 & 0.05 & 12.4 \\

Edge (Sobel) mask
& 0.455 & 0.967 & 0.049 & 0.04 & 12.4 \\

LiBRA (frequency mask)
& 0.458 & 0.970 & 0.046 & 0.03 & 12.4 \\
\midrule
$+\,L_{\mathrm{bit}}$
& 0.447 & 0.966 & 0.050 & 0.04 & 12.9 \\

Baseline, early-stopped
& 0.412 & 0.963 & 0.053 & 0.11 & 4.3 \\

$+\,L_{\mathrm{dfl}}$
& 0.456 & 0.958 & 0.058 & 0.03 & 13.6 \\
\bottomrule
\end{tabular}%
}
\end{table}

The spatial prior is not required for evasion, but it improves fidelity. Full-latent LiBRA and the frequency prior both reach a two-sided rate of $0.03$, while the prior raises SSIM from $0.962$ to $0.970$ and lowers LPIPS from $0.053$ to $0.046$ at essentially unchanged runtime. Random and Sobel masks yield SSIM and LPIPS values between those of the full-latent and frequency-guided configurations, with detection rates of $0.05$ and $0.04$, respectively. The frequency-guided mask provides the highest SSIM and lowest LPIPS among the four mask configurations, while matching the full-latent detection rate of $0.03$. Early stopping alone does not reproduce LiBRA: the early-stopped baseline reaches Bit ACC $0.412$ but leaves $11\%$ of images detectable, versus $3\%$ for LiBRA, and has lower SSIM ($0.963$ versus $0.970$) and higher LPIPS ($0.053$ versus $0.046$). Adding $L_{\mathrm{bit}}$ likewise does not help, slightly worsening both fidelity and detection rate.

\subsubsection{Negative result: blind spectral destruction}
As an auxiliary ablation, we test whether an additional spectral-disruption term improves LiBRA. The term measures Fourier-amplitude differences over the high-frequency subbands:
\begin{equation}
L_{\mathrm{dfl}}=\sum_{b\in\{LH,HL,HH\}}\big\lVert\,\lvert\mathcal{F}(b')\rvert-\lvert\mathcal{F}(b_0)\rvert\,\big\rVert_1,
\label{eq:dfl}
\end{equation}
We maximize this term to disrupt spectral amplitudes. This auxiliary experiment examines a spectral-disruption strategy studied for attacks without detector feedback~\cite{kassis2025unmarker}; the term is not part of LiBRA. Adding $L_{\mathrm{dfl}}$ leaves the two-sided detection rate unchanged at $0.03$, while lowering SSIM from $0.970$ to $0.958$, increasing LPIPS from $0.046$ to $0.058$, and increasing runtime from $12.4$ to $13.6$~s per image. In this comparison, the additional spectral term increases distortion and computation without improving evasion.

\subsection{Sensitivity to the Target Operating Point}\label{sec:exp_sens}
\begin{figure*}[!t]
    \centering
    \includegraphics[width=\textwidth]{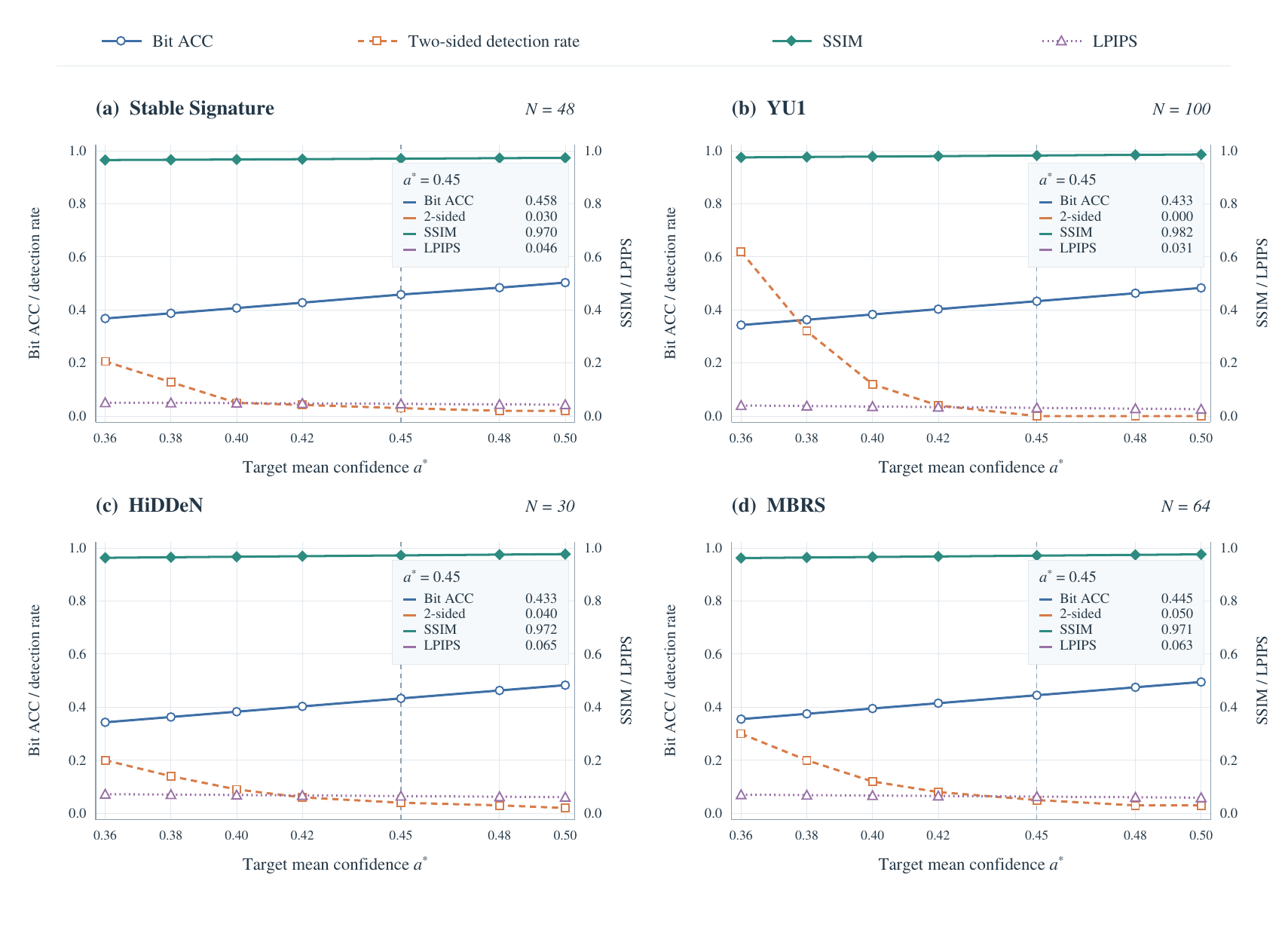}
   \caption{Sensitivity of LiBRA to the target mean confidence
$a^{\star}$ across four watermarking systems. The vertical
dashed line marks the default setting $a^{\star}=0.45$.}
    \label{fig:tradeoff}
\end{figure*}

We examine how the target mean confidence $a^{\star}$ affects hard bit accuracy, detection rate, and image fidelity on Stable Signature. \autoref{tab:sens} reports mean Bit ACC and the mean absolute error (MAE) between each image's hard bit accuracy and the target:
\[
\mathrm{MAE}=\frac{1}{5000}\sum_{j=1}^{5000}\left|a(x'_j)-a^{\star}\right|.
\]
This measures deviations across individual images. We use $a^{\star}=0.45$ as the common default throughout the main experiments and vary it here to examine its effect. \autoref{fig:tradeoff} presents a sweep of the target mean confidence across the four watermarking systems.

The MAE ranges from $0.026$ to $0.031$ across the tested targets. At $a^{\star}=0.30$, mean Bit ACC is $0.309$, below the lower acceptance boundary of $0.3125$, and the two-sided detection rate reaches $0.44$. Increasing the target from $0.40$ to $0.50$ reduces the detection rate from $0.05$ to $0.02$, raises SSIM from $0.967$ to $0.973$, and lowers LPIPS from $0.049$ to $0.043$. Thus, targets closer to $0.5$ improve both evasion and fidelity in this sweep. The default target of $0.45$ gives the lowest MAE, while $0.50$ gives the highest SSIM, lowest LPIPS, and a detection rate tied with $0.48$.

\begin{table}[t!]
\centering
\footnotesize
\setlength{\tabcolsep}{4pt}
\caption{Sensitivity to the target confidence.}
\label{tab:sens}
\begin{tabular}{lccccc}
\toprule
$a^{\star}$ & Bit ACC & \shortstack{MAE to\\target} & SSIM$\uparrow$ & LPIPS$\downarrow$ & $\mathrm{DR}_2\downarrow$ \\
\midrule
0.30 & 0.309 & 0.031 & 0.962 & 0.053 & 0.44 \\
0.40 & 0.407 & 0.027 & 0.967 & 0.049 & 0.05 \\
0.45 & 0.458 & 0.026 & 0.970 & 0.046 & 0.03 \\
0.48 & 0.484 & 0.027 & 0.972 & 0.044 & 0.02 \\
0.50 & 0.503 & 0.029 & 0.973 & 0.043 & 0.02 \\
\bottomrule
\end{tabular}
\end{table}

\subsection{Point Target versus Band Penalty}\label{sec:exp_band}
We compare LiBRA's point-target objective with a penalty that allows mean soft confidence to lie within a band around $0.5$:
\begin{equation}
L_{\mathrm{band}}=\operatorname{ReLU}\!\big(\lvert\bar c-0.5\rvert-h\big)^{2},
\label{eq:bandloss}
\end{equation}
The band penalty is zero when $|\bar c-0.5|\le h$, allowing a range of mean soft confidences instead of penalizing deviations from a single target. \autoref{tab:bandloss} compares the two objectives on Stable Signature at $N=48$ and $\alpha=0.01$. The hard-test acceptance set has boundary accuracies of $0.3125$ and $0.6875$, corresponding to a half-width of $0.1875$ around $0.5$. We evaluate both objectives using the resulting hard-bit detection decisions and image fidelity. The $\mathrm{DR}_2$ column reports detection before post-processing, while ``DR after JPEG'' reports the two-sided detection rate after JPEG compression at quality factor $q=50$.

The band penalty exposes a direct margin--fidelity trade-off rather than strictly dominating the point target. With $h=0.10$, it improves SSIM from $0.970$ to $0.972$ and LPIPS from $0.046$ to $0.044$, but its clean two-sided rate increases from $0.03$ to $0.05$; after JPEG the corresponding rates are $0.05$ and $0.06$. Wider bands save slightly more distortion but lose substantially more evasion: at $h=0.15$ and $0.18$, the clean two-sided rates rise to $0.12$ and $0.26$. We therefore retain the point target as the more conservative operating point in these experiments: its small fidelity cost buys a measurable margin in both the clean and post-processed evaluations.

\begin{table}[t!]
\centering
\footnotesize
\setlength{\tabcolsep}{4pt}
\caption{Comparison of point-target and band-penalty objectives.}
\label{tab:bandloss}
\begin{tabular}{lccccc}
\toprule
Objective & Bit ACC & SSIM$\uparrow$ & LPIPS$\downarrow$ & $\mathrm{DR}_2\downarrow$ & \shortstack{DR after\\JPEG$\downarrow$} \\
\midrule
Point, $a^{\star}=0.45$ & 0.458 & 0.970 & 0.046 & 0.03 & 0.05 \\
Band, $h=0.10$          & 0.406 & 0.972 & 0.044 & 0.05 & 0.06 \\
Band, $h=0.15$          & 0.359 & 0.974 & 0.042 & 0.12 & 0.09 \\
Band, $h=0.18$          & 0.331 & 0.975 & 0.041 & 0.26 & 0.15 \\
\bottomrule
\end{tabular}
\end{table}

\subsection{Sensitivity to the Defender's Significance Level}\label{sec:exp_alpha}
The exact acceptance set depends on two quantities: the key length $N$, which the attacker knows, and the significance level $\alpha$, which it does not. \autoref{sec:setup} gives the exact accepted count ranges for the evaluated key lengths at $\alpha=0.01$; here we use YU1 with $N=100$ and vary $\alpha$. Acceptance bounds are computed using exact binomial tests. ``Width'' is the difference between the largest and smallest accepted bit accuracies. Under the independent fair-bit null, $\alpha$ is an upper bound on the false-positive rate. No re-optimization is required, because the attacked images and their decoded bits are fixed and changing $\alpha$ only moves the threshold of the test applied to them: we re-evaluate the same $5{,}000$ images under detectors that use different $\alpha$.

\autoref{tab:alpha} shows that decreasing $\alpha$ widens the exact acceptance set. LiBRA has zero observed detections at both $\alpha=0.01$ and $\alpha=0.001$. Increasing $\alpha$ narrows the acceptance set and raises the detection rate to $0.06$ at $\alpha=0.05$ and $0.21$ at $\alpha=0.1$. The provider can therefore detect more attacked images by enlarging the rejection region, at the cost of a higher false-positive rate under the stated null model. Here, $\alpha$ specifies the test's nominal significance level and bounds its false-positive rate; the exact rate depends on the discrete binomial distribution.

\begin{table}[t!]
\centering
\footnotesize
\setlength{\tabcolsep}{4pt}
\caption{Effect of the detector significance level.}
\label{tab:alpha}
\begin{tabular}{lccccc}
\toprule
& & & \multicolumn{3}{c}{$\mathrm{DR}_2\downarrow$} \\ \cmidrule(lr){4-6} $\alpha$ & \shortstack{Acceptance\\bounds} & Width & Baseline & B+ & LiBRA \\
\midrule
$0.1$ & $[0.420,0.580]$ & 0.160 & 1.00 & 1.00 & 0.21 \\
$0.05$ & $[0.400,0.600]$ & 0.200 & 1.00 & 1.00 & 0.06 \\
$0.01$ & $[0.370,0.630]$ & 0.260 & 1.00 & 0.98 & 0.00 \\
$0.001$ & $[0.340,0.660]$ & 0.320 & 1.00 & 0.93 & 0.00 \\
\bottomrule
\end{tabular}
\end{table}

\subsection{Relaxing Knowledge of the Key}\label{sec:exp_key}

Our threat model assumes that the adversary knows the secret key $w$. We also evaluate LiBRA on Stable Signature using an estimated key $\hat w$ that differs from $w$ in a fraction $\epsilon$ of its bit positions. The attack optimizes against the estimated key $\hat w$, while all metrics are computed against the true key $w$ used by the detector.

\autoref{tab:keyerr} shows the effect of key estimation errors on LiBRA. As $\epsilon$ increases from $0$ to $0.20$, mean Bit ACC rises from $0.458$ to $0.481$, and the MAE to the target, defined in \autoref{sec:exp_sens}, increases from $0.026$ to $0.041$. The two-sided detection rate increases from $0.03$ to $0.05$; SSIM and LPIPS change from $0.970/0.046$ to $0.968/0.048$. Across the tested key error rates up to $20\%$, LiBRA maintains low detection rates and similar image fidelity.

\begin{table}[t!]
\centering
\footnotesize
\setlength{\tabcolsep}{4pt}
\caption{Sensitivity to watermark key estimation errors.}
\label{tab:keyerr}
\begin{tabular}{lccccc}
\toprule
Key error $\epsilon$ & Bit ACC & \shortstack{MAE to\\target} & SSIM$\uparrow$ & LPIPS$\downarrow$ & $\mathrm{DR}_2\downarrow$ \\
\midrule
$0$ (exact key) & 0.458 & 0.026 & 0.970 & 0.046 & 0.03 \\
$0.05$          & 0.463 & 0.028 & 0.970 & 0.046 & 0.03 \\
$0.10$          & 0.469 & 0.032 & 0.969 & 0.047 & 0.04 \\
$0.20$          & 0.481 & 0.041 & 0.968 & 0.048 & 0.05 \\
\bottomrule
\end{tabular}
\end{table}

\subsection{Effect of the Iteration Budget}\label{sec:exp_iters}
\autoref{prop:base} states that, in margin space, the baseline objective has no finite stationary point, whereas the two-sided objective has a zero set (\autoref{prop:twosided}). We compare the baseline and LiBRA on Stable Signature while varying the iteration budget $T$ to examine their optimization behavior.

\autoref{tab:iters} shows different responses to increasing the iteration budget. For the baseline, mean Bit ACC decreases from $0.214$ at $T=10$ to $0.004$ at $T=100$, while SSIM falls from $0.968$ to $0.921$. Its two-sided detection rate reaches $1.00$ at $T=25$ and remains unchanged as optimization continues. LiBRA already achieves a detection rate of $0.06$ at $T=10$, which decreases to $0.03$ at $T=25$. From $T=25$ to $T=100$, its detection rate stays at $0.03$, SSIM remains between $0.970$ and $0.971$, and mean Bit ACC changes from $0.466$ to $0.455$. These results show that further iterations continue to degrade the baseline's image fidelity, while LiBRA maintains stable evasion and fidelity over the tested budgets from $25$ to $100$.
\begin{table}[t!]
\centering
\footnotesize
\setlength{\tabcolsep}{4pt}
\caption{Effect of the iteration budget.}
\label{tab:iters}
\begin{tabular}{lcccccc}
\toprule
& \multicolumn{3}{c}{Baseline} & \multicolumn{3}{c}{LiBRA} \\
\cmidrule(lr){2-4}\cmidrule(lr){5-7}
$T$ & Bit ACC & SSIM$\uparrow$ & $\mathrm{DR}_2\downarrow$ & Bit ACC & SSIM$\uparrow$ & $\mathrm{DR}_2\downarrow$ \\
\midrule
$10$  & 0.214 & 0.968 & 0.94 & 0.527 & 0.975 & 0.06 \\
$25$  & 0.052 & 0.955 & 1.00 & 0.466 & 0.971 & 0.03 \\
$50$  & 0.018 & 0.942 & 1.00 & 0.458 & 0.970 & 0.03 \\
$100$ & 0.004 & 0.921 & 1.00 & 0.455 & 0.970 & 0.03 \\
\bottomrule
\end{tabular}
\end{table}

\subsection{Effect of Image Resolution}\label{sec:exp_res}
We measure how image resolution affects LiBRA's detection outcomes, per-image runtime, and peak memory usage.

\autoref{tab:res} compares LiBRA on Stable Signature images at three resolutions with $T=50$. The autoencoder downsamples each spatial dimension by a factor of $8$. Mean Bit ACC remains between $0.457$ and $0.461$, and two-sided detection rates remain between $0.03$ and $0.04$, showing similar evasion across the tested resolutions. Increasing resolution from $256\times256$ to $768\times768$ increases runtime from $3.6$ to $25.9$~s per image, approximately $7.2\times$, and peak memory from $5.1$ to $29.6$~GB.

\begin{table*}[t!]
\centering
\footnotesize
\setlength{\tabcolsep}{4pt}
\caption{Effect of image resolution.}
\label{tab:res}
\begin{tabular}{lccccc}
\toprule
Resolution & \shortstack{Latent shape\\$(H/f)\times(W/f)\times C$} & Bit ACC & $\mathrm{DR}_2\downarrow$ & Time (s/image) & \shortstack{Peak memory\\(GB)} \\
\midrule
$256\times256$ & $32\times32\times4$ & 0.461 & 0.04 & 3.6 & 5.1 \\
$512\times512$ & $64\times64\times4$ & 0.458 & 0.03 & 12.4 & 13.8 \\
$768\times768$ & $96\times96\times4$ & 0.457 & 0.03 & 25.9 & 29.6 \\
\bottomrule
\end{tabular}
\end{table*}

\subsection{Robustness of the Attacked Image}\label{sec:exp_survive}
An attacked image is rarely consumed in the exact form the attacker produces: it is recompressed when uploaded, resized when displayed, possibly screenshotted. Although LiBRA targets an interior soft-confidence value, subsequent image processing may shift the hard bit accuracy outside the detector's acceptance set. We therefore examine
whether the achieved evasion survives common post-processing
operations. \autoref{tab:survive} reports results on $5{,}000$ Stable Signature images. The upper part evaluates detection after standard post-processing of images attacked with LiBRA. The lower part reports the combined detection rate under the stronger detector discussed in \autoref{sec:discussion}, which additionally tests the dispersion of per-bit confidences. Each statistical test uses $\alpha=0.01$.

LiBRA retains low two-sided detection rates after the tested post-processing transformations. Starting from $0.03$ before post-processing, the detection rate remains between $0.03$ and $0.05$ after JPEG compression, resizing, Gaussian noise, and center cropping. Image fidelity changes with the transformation, with SSIM decreasing to $0.611$ after cropping. The dispersion-aware detector produces a different result: its combined detection rate is $0.64$ for LiBRA and $1.00$ for the no-attack, baseline, and B+ settings. Confidence dispersion therefore provides additional detection information beyond hard bit agreement.

\begin{table}[t!]
\centering
\footnotesize
\setlength{\tabcolsep}{2pt}
\caption{Post-processing robustness and dispersion-aware detection.}
\label{tab:survive}
\begin{tabular}{lcccc}
\toprule
Condition & Bit ACC & SSIM$\uparrow$ & $\mathrm{DR}_1\downarrow$ & $\mathrm{DR}_2\downarrow$ \\
\midrule
None (as attacked)             & 0.458 & 0.970 & 0.00 & 0.03 \\
JPEG ($q=90$)                  & 0.466 & 0.962 & 0.12 & 0.03 \\
JPEG ($q=50$)                  & 0.489 & 0.921 & 0.36 & 0.05 \\
Resize ($0.5\times$, restored) & 0.474 & 0.931 & 0.21 & 0.04 \\
Gaussian noise ($\sigma=0.02$) & 0.471 & 0.903 & 0.18 & 0.04 \\
Center crop ($10\%$)           & 0.483 & 0.611 & 0.29 & 0.05 \\
\midrule
\multicolumn{5}{l}{Dispersion-aware detection} \\
\midrule
Setting & No attack & Baseline & B+ & LiBRA \\
Combined DR$\downarrow$ & 1.00 & 1.00 & 1.00 & 0.64 \\
\bottomrule
\end{tabular}
\end{table}

\subsection{An Adaptive Round Against the Dispersion-Aware Detector}\label{sec:exp_adaptive}
The detector in \autoref{sec:exp_survive} combines hard-bit agreement with a test of per-bit confidence dispersion. We evaluate an adaptive LiBRA variant that accounts for this additional statistic during optimization.

The adaptive objective adds a penalty that guides confidence variance toward a reference value estimated from unwatermarked images:
\begin{equation}
L_{\mathrm{disp}}=\Big(\operatorname{Var}_i\big(\sigma(d_i w_i)\big)-v^{\star}\Big)^{2},
\label{eq:disp}
\end{equation}
where $v^{\star}$ is estimated once from a batch of unwatermarked images. Adding $\lambda_{\mathrm{disp}}L_{\mathrm{disp}}$ to \autoref{eq:total} costs one extra scalar per iteration and no additional forward pass, since $d$ is already computed.

\autoref{tab:adaptive} compares the unaware and adaptive attacks on Stable Signature, with $\alpha=0.01$ for each test. $\mathrm{DR}_2$, ``Dispersion DR'', and ``Combined DR'' report detection rates for the hard-bit test, the dispersion test, and the combined detector, respectively. The unaware attack optimizes \autoref{eq:total}, while the adaptive attack adds the dispersion penalty in \autoref{eq:disp}.

\begin{table}[t!]
\centering
\footnotesize
\setlength{\tabcolsep}{3pt}
\renewcommand{\arraystretch}{1.1}
\caption{Adaptive attack against the dispersion-aware detector.}
\label{tab:adaptive}
\resizebox{\columnwidth}{!}{%
\begin{tabular}{@{}lccccc@{}}
\toprule
Setting
& $\mathrm{DR}_2\downarrow$
& \shortstack{Dispersion\\DR$\downarrow$}
& \shortstack{Combined\\DR$\downarrow$}
& SSIM$\uparrow$
& LPIPS$\downarrow$ \\
\midrule
No attack
& 1.00 & 0.99 & 1.00 & 1.000 & 0.000 \\

LiBRA
& 0.03 & 0.62 & 0.64 & 0.970 & 0.046 \\

\textbf{LiBRA $+L_{\mathrm{disp}}$}
& 0.04 & 0.06 & 0.09 & 0.966 & 0.050 \\
\bottomrule
\end{tabular}%
}
\end{table}

 Adding $L_{\mathrm{disp}}$ reduces the combined detection rate from $0.64$ to $0.09$ and the dispersion-only detection rate from $0.62$ to $0.06$. The original two-sided detection rate changes from $0.03$ to $0.04$, while SSIM decreases from $0.970$ to $0.966$ and LPIPS increases from $0.046$ to $0.050$. These results show that LiBRA can accommodate the additional detection statistic through a variance-matching term, substantially reducing combined detection with a small change in the measured fidelity.

\subsection{Qualitative Results}\label{sec:exp_qual}
\begin{figure*}[!t]
    \centering
    \includegraphics[width=0.90\textwidth,trim=0 0 0 55,clip]
        {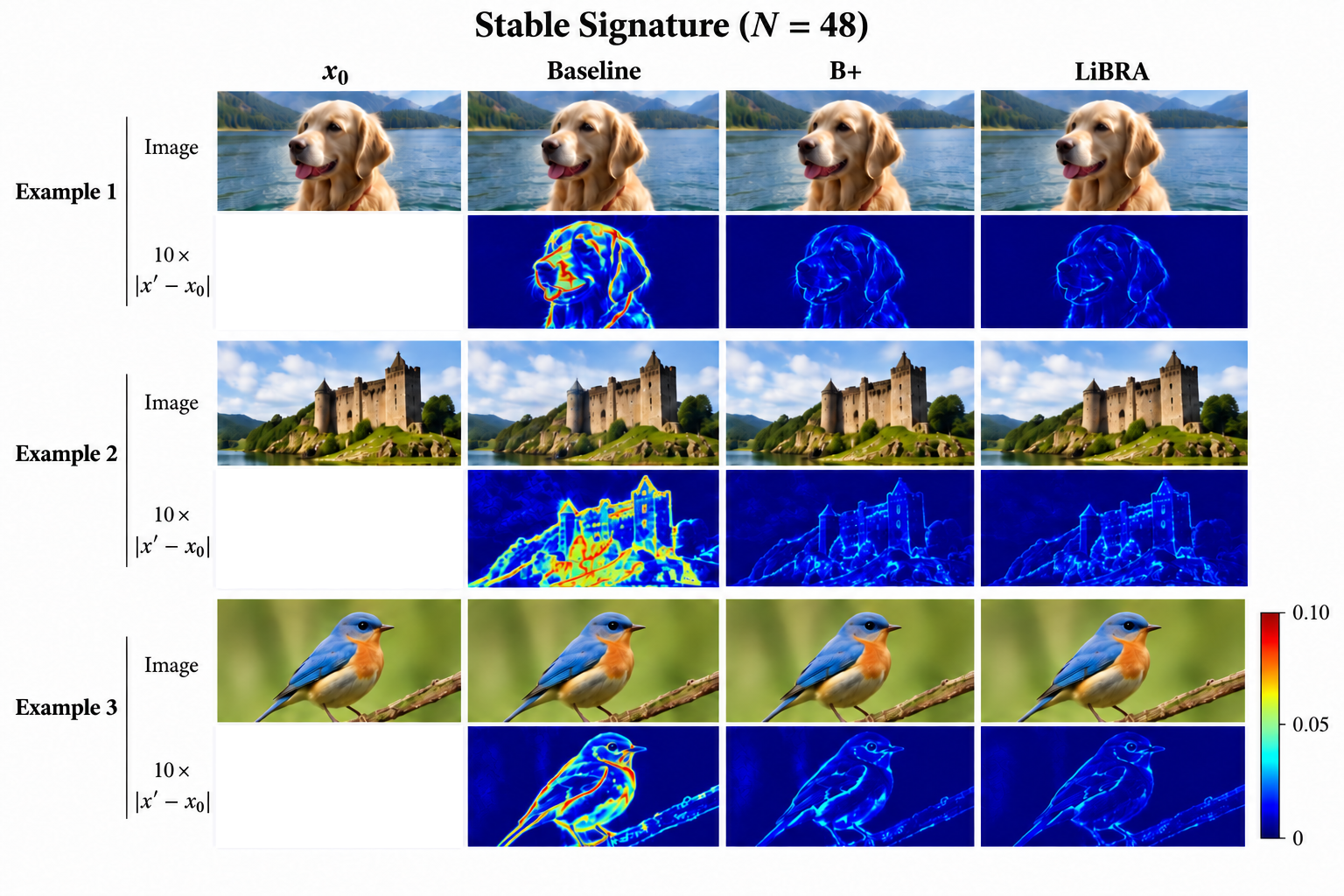}
    \caption{Qualitative comparison on Stable Signature ($N=48$). Columns: input $x_0$, baseline, B+, and LiBRA; lower rows are labeled as amplified residuals.}
    \label{fig:qual}
\end{figure*}

\autoref{fig:qual} compares three examples from Stable Signature. The first column shows the watermarked input $x_0$, followed by the outputs of the baseline, B+, and LiBRA. Beneath each attacked image, the residual map visualizes its difference from the corresponding input. Residual maps are omitted for the input column because its difference from itself is zero.

\section{Discussion}\label{sec:discussion}

\subsection{Evaluation Must Match the Decision Rule}
The same attacked images can count as complete successes under the naive one-sided rule and complete failures under the two-sided test. An inverted decoding retains information about the key, so reporting low bit accuracy alone obscures attribution risk. Evaluation should specify the tested statistic, null model, significance level, and image-level detection rate. Comparing level-calibrated detectors is also necessary when making claims about security at matched false-positive rates; our naive one-sided comparison does not provide that calibration.

\subsection{Implications for Watermark Design}
The four evaluated systems share a soft per-bit decoding interface despite different embedding mechanisms. Our objective controls only the mean of those confidences. Dispersion, joint bit patterns, or agreement across disjoint key subsets may expose information that this mean leaves unconstrained. \autoref{sec:exp_survive} and~\autoref{sec:exp_adaptive} examine dispersion and show why the resulting gain must be assessed against an aware attacker. Redundant keys and joint-pattern tests remain future directions. Multiple rejection rules also require calibration of their combined false-positive rate, for example using a multiple-testing correction~\cite{holm1979}; a level of $0.01$ per test does not imply the same overall level.

\subsection{Bidirectional Control and Detection Outcomes}
LiBRA separates the direction encouraged by the optimization objective from the final detection decision. \autoref{prop:hinge} shows that per-bit thresholding can still admit fully inverted decodings, whereas \autoref{prop:twosided} shows that LiBRA reverses its gradient direction when mean confidence crosses the target. This provides a mechanism for counteracting excessive inversion while retaining flexibility in individual bit configurations. Reaching the confidence target does not, however, guarantee hard-test acceptance. We therefore evaluate final detection outcomes alongside image fidelity and use controlled comparisons to distinguish the effects of objective design, optimization space, and key knowledge.

\subsection{Limitations and Threats to Validity}
The attack assumes detector gradients and access to the key or an estimate. It therefore characterizes the stated white-box setting, not an arbitrary black-box deployment. The key-error experiment relaxes one assumption but does not supply missing detector gradients. The target also depends on an assumed significance level; different null distributions, statistics, or changing thresholds can require a different objective.

The evaluation covers four released checkpoints and their associated data, rather than a broad sweep of training runs. It does not establish transfer to initial-noise schemes such as Tree-Ring and Gaussian Shading~\cite{wen2023treering,yang2024gaussian}, whose inversion-based detectors differ from the evaluated interfaces. The spatial prior uses one configuration; interactions between its thresholds and the watermark objective remain unexplored.

The reproduced baseline additionally retains a sign-consistency term that LiBRA does not use. The ablations and early-stopping control help separate these factors, but no single comparison establishes universal superiority. Finally, reaching a mean-confidence target neither guarantees hard-count evasion nor minimizes distortion. Per-image outcomes and uncertainty estimates are needed to interpret aggregate means and observed zero detection rates.

\subsection{Ethical Considerations}
Watermark-removal research can improve robustness evaluation and can also enable provenance evasion. The study is scoped to publicly released research implementations and controlled evaluation images under explicit capability assumptions. The supplied manuscript reports no targeting of commercial provenance services, private APIs, or deployed infrastructure. Our defensive analysis therefore accompanies the attack, including the adaptive failure of a stronger detector; it should not be read as a claim that adding dispersion alone resolves the vulnerability.

\section{Conclusion}\label{sec:conclusion}
We present LiBRA, a watermark removal method based on bidirectional latent optimization. Its symmetric mean-confidence objective counteracts excessive inversion while allowing fidelity constraints to favor less damaging changes. Experiments across four released watermarking systems show improved evasion and image quality over the inversion baseline. An iteration study on Stable Signature further shows stable evasion and fidelity from $25$ to $100$ iterations. Our analysis also confirms that reaching the soft target does not guarantee successful removal. Future work will explore black-box attacks, additional detection statistics, and minimum-distortion limits.

\bibliographystyle{IEEEtran}
\bibliography{references}

\end{document}